# Stark Effect of Doped Two-Dimensional Transition Metal Dichalcogenides

Xiaobo Lu[1] and Li Yang[1, 2*]

[1]Department of Physics and Institute of Materials Science and Engineering, Washington University, St. Louis, Missouri, 63130, USA

[2]Institute of Materials Science and Engineering, Washington University in St. Louis, St. Louis, Missouri 63130, USA

**Email**: lyang@physics.wustl.edu

**Abstract:** The band gap of two-dimensional (2D) semiconductors can be efficiently tuned by gate electric field, which is so called the Stark effect. We report that doping, which is essential in realistic devices, will substantially change the Stark effect of few-layer transition metal dichalcogenides in unexpected ways. Particularly in bilayer structures, because of the competition between strong quantum confinement and intrinsic screening length, electron and hole dopings exhibit surprisingly different Stark effects: doped electrons actively screen the external field and result in a nonlinear Stark effect; however, doped holes do not effectively screen the external field, causing a linear Stark effect that is the same as that of undoped materials. Our further analysis shows that this unusual doping effect is not limited within transition metal dichalcogenides but general for 2D structures. Therefore, doping plays a much more crucial role in functional 2D devices and this unusual Stark effect also provides a new degree of freedom to tune band gaps and optical properties of 2D materials.

Band gap may be the most fundamental, important character to decide properties and applications of materials. Realizing a tunable band gap is thus highly desired. Recently emerged layered two-dimensional (2D) semiconductors shed light on new opportunities to physically tune band gaps [1-7]. Theoretical calculations and measurements have shown that band gaps of few-layer transition-metal dichalcogenides (TMDs) [8-11] and nano-flake black phosphorus (BP) [12-16] can be reduced by a few hundred meV via applying a gate (off-plane) field, which is so called the Stark effect of 2D structures [8,15,17]. This approach is particularly intriguing for applications because gate structures [18-20] can be seamlessly integrated into modern microelectronic devices for dynamically tuning the band gap.

To date most research on this field has focused on tunable band gaps of intrinsic structures. However, doping is essential for realizing functional devices [21]. Those doped or thermally excited free carriers will inevitably enhance the electric screening and resist the applied gate field [22, 23]. In contrast to bulk materials, free carriers in few-layer structures will be spatially accumulated or attenuated around boundaries (surfaces) by external field, resulting in an inhomogeneous polarization and impacting the tunable band gap. Particularly, if the thickness of these ultra-thin structures approaches the characteristic screening length, quantum effects may become significant to affect the charge redistribution, giving rise to unusual screening effects and novel doping effects.

In this work, we show that ultra-thin boundary conditions and corresponding enhanced quantum confinement will significantly impact the 2D Stark effect. Take 2H-phase $MoS_2$ as an example, which is a typical 2D semiconductor intensively studied. We find that its bilayer structure exhibits a pronounced difference between electron and hole dopings: electron doping induces a nonlinear dependence of the band gap on the gate field while hole doping induces a linear band gap reduction. Interestingly, for those thicker structures with a layer number more than two, both electron and hole dopings exhibit the same nonlinear band gap reduction under the gate field. Our first-principles calculations and model show that this unusual Stark effect is essentially from the competition between the characteristic screening length and quantum confinement of 2D structures. This study predicts that the tunable band gap in realistic 2D devices is much more complicated than what have been expected. Meanwhile, the doping effect opens the door for designing functional devices with nonlinear/linear tunable band gaps.

The atomic structure of few-layer $MoS_2$ is shown in figure 1 (a). We employ the density functional theory (DFT) with the Perdew-Burke-Ernzerh(PBE) exchange-correlation potential and include the van del Waals (vdW) correction [24-26]. Doping is introduced by the rigid-band doping approach [27], which can avoid the large supercell while still captures essentially electrostatic doping effects. To describe realistic doping densities, e.g., 0.001 electron/unit cell, we apply a non-uniform k-point sampling scheme, in which the k-grid is particularly enhanced around the Fermi surface in the reciprocal space (around band edges in this study). Consequently, we estimate the error bar of our calculation is also decided by the energy resolution of the k-point sampling. The force and stress are fully relaxed under different layer thickness and doping densities. It must be pointed out that we confine the displacement field magnitude within 1 V/nm, which is well within the practical device range [1,10,12].

*Bilayer results:* We start from 2H-phase bilayer $MoS_2$ with the A-A' stacking style, which has the lowest total energy [9, 28]. The DFT calculated pristine band structure is presented in figure 1 (b). As expected, the band gap is indirect (~ 1.2 eV) because of the interlayer interaction [29, 30]. An important character is that the band hybridization of the conduction band minimum (CBM) at the K (K') point is much smaller than that of the valence band maximum (VBM) at the Γ point; the VBM splitting is large (~ 577 meV) while the CBM splitting (~ 1 meV) is small, as magnified in figure 1 (c) (the black line).

Numerous works have shown that the band gap of bilayer $MoS_2$ can be efficiently reduced by applying gate field [8-11], which is known as the Stark effect in 2D materials. To understand the origin of this band-gap reduction, we plot the band edge splittings under gate field (D=1.0V/nm) in figure 1 (c) (the blue lines). The CBM splitting is substantially enlarged while the VBM splitting is nearly unchanged, indicating that the band gap reduction is mainly from the splitting of the conduction band edge. Moreover, as presented in figure 2 (a), the band gap of the intrinsic bilayer $MoS_2$ (the red line) is linearly reduced by the gate field [8,10]. This linear band-gap reduction is from the fact that gate field generates a potential difference between layers and creates a band offset. The offset is thus approximately decided by the net electrical potential drop between two layers, which is roughly about the bare potential drop ($\Delta \varphi = E \cdot d$, where $E$ is the external gate field and $d$ is the effective interlayer distance) screened by the vertical dielectric constant. Therefore, the reduction of the overall band gap is $\Delta E_g \propto E \cdot d/\varepsilon_\perp$. Based on this picture, the vertical dielectric constant ($\varepsilon_\perp$), i.e., the vertical screening effect, is crucial for deciding the band gap variation [31]. This is the motivation of that doping may substantially impact the Stark effect.

*Electron doping:* We first focus on the tunable band gap of electron-doped (*n*-type) bilayer $MoS_2$, as summarized in figure 2 (a). Contrast to the linear reduction of band gap of the intrinsic case, the pronounced feature of electron-doped bilayer $MoS_2$ is the nonlinear reduction by gate field. Initially, the band gap is nearly fixed under a weak gate field. This is due to those doped free carriers that effectively screen the external field, reflecting the metallic nature of doped bilayer MoS$_2$. However, above a threshold field, the band gap starts to reduce linearly with the same slope as that of the intrinsic bilayer MoS$_2$, indicating that doped free carriers no longer screen the stronger external field. In other words, doped bilayer MoS$_2$ exhibits two "quantized" screenings, the metallic one under weak gate field and the semiconducting one under strong gate field.

The first-principle calculated electronic band structures and wavefunctions provide us the explanation of these nonlinear band-gap variations under gate field. Before applying gate field, the lowest two conduction bands of bilayer MoS$_2$ are nearly degenerated at the CBM and those doped carriers are thus evenly distributed in both layers. Take an example of a doping density of 0.4% (around $4.6 \times 10^{12} cm^{-2}$). As shown in figure 2 (b), under a weak gate field (0.2 V/nm), the degeneracy is slightly broken. With the same Fermi level, the number of occupied states of these two bands are different. This spatial redistribution of charge density and subsequent polarization screen the applied gate field, contributing to a metallic screening and a nearly unchanged band gap under the weak gate field. When the electric field reaches a critical value (0.4 V/nm), as shown in figure 2 (c), the energy splitting is large enough that the Fermi level starts to cross only one of these two conduction bands. In this case, because all doped carriers are distributed in the other layer, there is no vertically mobile carrier and the vertical screening is not effective any more, making the screening of doped bilayer structure transit from metal to intrinsic semiconductor. For higher gate field (0.8 V /nm) shown in figure 2 (d), the CBM splitting is even larger and the Fermi level still only crosses in the lowest conduction band and no longer effectively screens gate field.

The above discussion is supported by the plot of the variation of charge density of free electrons in figure 2 (e), in which the carrier density is plotted along the z direction with integration of the x-y plane (the definition of x, y, and z directions are shown in figure 1 (a)). Under a strong gate field (~ 0.8 V/nm), most of doped electrons (~ 77%) are transferred into the lower-energy conduction band (L1). This mechanism also explains the density dependence of the critical field, which divides the flat and sloping regions in figure 2 (a): a higher doping density corresponds to a stronger critical gate field. This is because a higher doping density needs a higher critical gate field to create a larger splitting of CBM and to exhaust those free carriers from one layer to the other. Finally, noticeable fluctuations (~ a few meV) of band gap can be observed in our first-principles results, such as figure 2 (a). This is due to the artefact of DFT calculations.

*Hole doping:* Then we turn to the case of hole-doped (p-type) bilayer MoS$_2$. As shown in figure 3 (a), unlike the nonlinear behavior of electron-doped cases, the band gap reduction of hole-doped bilayer MoS$_2$ is surprisingly linear and it is similar to that of intrinsic bilayer MoS$_2$. To understand this, we have plotted charge densities of doped holes. In figure 3 (b), without the significant charge transfer observed electron-doped cases, the hole density is nearly frozen even under a strong field. A strong gate field (1.0 V/nm) only creates a 4% charge transfer between two layers. Therefore, the polarization and corresponding screening are weak, giving rise to the nearly same band-gap reduction as that of intrinsic bilayer $MoS_2$.

This "frozen" hole density and screening are essentially from the quantum confinement and strong interlayer hybridization of hole states. In ultra-thin structures, free carriers may not move freely along the off-plane direction because they need available quantum states. As shown in figure 1 (c), the VBMs of two layers are strongly hybridized because of antibonding nature of the states that are mainly from the delocalized Mo $d_{z^2}$ and sulfur $p_z$ orbitals. [8] Two consequences are thus expected: 1) a larger energy splitting of VBM is created; 2) the wavefunctions of each of two bands are evenly distributed in both layers. Typical gate field cannot change energies of these two bands and their hybridization, as evidenced in figure 1 (c). A tight-binding model in supplementary information further confirms this depressed charge transfer in bilayer $MoS_2$ and show that the strong interlayer coupling is the main reason. Therefore, all available hole states are nearly evenly distributed in both gated layers. Charge transfer between two layers is limited and the screening effect is substantially depressed. From the point view of electrodynamics, the characteristic screening length of holes in $MoS_2$ is, at least, not less than the thickness of two atomic layers.

*Multilayer results:* For thicker doped $MoS_2$ with a layer number more than 2, their general features of the tunable band gap (Stark effect) are similar. More details can be found in supplementary information. Therefore, we focus on the case of trilayer $MoS_2$, whose band gap variations are summarized in figures 4 (a) and (c). Surprisingly, unlike the bilayer case, in which electron and hole dopings show substantially different Stark effects, those of trilayer $MoS_2$ exhibit a similarly nonlinear evolution of the band gap under gate field: they are metallic under weak gate field and semiconducting under strong gate field. It is noticeable that the initial band gap of hole-doped trilayer $MoS_2$ is more sensitive to the hole doping density: a variation of 20 meV band gap variation is observed in figure 4 (c). This is a quantum confinement effect: the doped holes form an inhomogeneous charge distribution, which induces an extra electrical potential well between surface and middle layers. As a result, the localized conduction band edge is split and, particularly, the conduction band of the middle layer will be substantially lowered. This will reduce the overall band gap. On the other hand, for thicker $MoS_2$ samples, the quantum confinement effect is weaker, making this hole-doping induced band gap variation smaller. More details can be found in supplementary information.

For the electron-doped cases, the reason of this nonlinear evolution of the band gap is similar with that of bilayer, which is due to the one-atomic-layer screening length. As shown in figure 4 (b), the carrier charge distribution is efficiently transferred and saturated into the boundary layer under strong electric field, e.g., 75% of electron carriers are accumulated in the boundary L1 layer under a 1.0 V/nm gate field. The different charge density in the other two layers (L2 and L3) is from the boundary effect of quantum confinement. For the hole-doped cases, we have plotted their charge distributions in figure 4 (d). Interestingly, unlike the hole-doped bilayer, which the charge transfer between layers is nearly frozen, we observe a significant charge transfer. Moreover, unlike electron-doped cases, in which electrons can be saturated into one layer by external field, those doped holes are finally saturated into two layers. These charge transfers indicate that the characteristic screening length of holes of $MoS_2$ is about the thickness of two atomic layers.

In fact, the characteristic screening length and the Stark effect are strongly correlated with interlayer interactions and band hybridization. In electron-doped $MoS_2$, the interlayer interaction is negligible, as evidenced by the tiny energy splitting of the CBM (figure 1 (c)). Therefore, the electron screening length is about the thickness of monolayer. For valence bands, the interlayer interaction is strong, as seen from the substantial energy splitting of VBM (figure 1 (c)). Moreover, the dominant interlayer interaction is the nearest-neighboring coupling. Therefore, the hole screening length is about the thickness of two atomic layers. Our tight-binding calculations presented in supplementary information support this relation between interlayer interactions and screening length. In this sense, we can expect different Stark effects in other few-layer 2D materials with different interlayer couplings. For example, in other few-layer TMDs in the 2H phase, because their interlayer interactions are similar to those of MoS2, their Stark effects under doping shall be similar to these results in this work. However, in black phosphorus, because of the similar interlayer couplings of conduction bands and valence bands, their band gap variation will be similar for both electron and hole dopings.

Finally, two extrinsic factors are neglected in this study, which may impact the above quantitative results. First, these DFT and tight-binding calculations are under the low-temperature condition. For higher temperature, the smearing of the Fermi level will smooth the transition between the metallic screening and semiconducting one although the basic features of flat and sloping reduction of band gaps shall still be robust. Particularly for the hole-doped bilayer $MoS_2$, the unique semiconducting screening will not be significantly affected because the strong interlayer coupling energy (~300 meV) is much larger than the typical room temperature thermal energy (~30 meV). Second, the quasiparticle band gap of 2D semiconductors can be substantially renormalized by doped free carriers. [32-34] The gate field will generate an inhomogeneous charge distribution, making many-electron effects different in different layers. Therefore, the values of tunable band gaps will be modified. However, the general features, such as different Stark effects of electrons and holes in bilayer $MoS_2$ and *etc.* shall not be qualitatively altered by these higher-order corrections.

In conclusion, we reveal the unexpected band gap variations in doped few-layer 2D semiconductors. Take the widely studied $MoS_2$ as an example, we show that the strong vertical quantum confinement and unique vdW interactions result in qualitatively different screening and Stark effects. Particularly, the screening length and the linearity of the Stark effect are essentially decided by the range of interlayer coupling. For weakly interlayer interacting states (electron-doping), doped free carriers can efficiently screen the gating field and make the band gap reduction very inefficient until all mobile carriers are accumulated into one surface layer. For strongly interacting states (hole-doping), depending on the interlayer interaction range, their screening can be frozen in bilayer structures while be efficient in thicker samples. Our study shows rich screening effects and Stark effects in doped few-layer vdW materials. This will be of fundamental importance for understanding current measurements and designing novel devices based on 2D semiconductors to realize linear/nonlinear tunable band gaps.

**Supplementary Material**
In the supplementary material we include the band gap variation of doped quadrilayer $MoS2$ under gate field. We also present a tight-binding model which gives the similar doping effects calculated by DFT. At the end of supplementary material, we discuss the small band-gap variation of doped multilayer $MoS2$ at zero field.

**ACKNOWLEDGMENT**
We acknowledge fruitful discussions with Vy Tran, Hongxia Zhong, and Shiyuan Gao. We are supported by the National Science Foundation (NSF) CAREER Grant No. DMR-1455346, NSF EFRI-2DARE-1542815, and the International Center for Advanced Renewable Energy & Sustainability (I-CARES). The computational resources have been provided by the Stampede of Teragrid at the Texas Advanced Computing Center (TACC). This work used the Extreme Science and Engineering Discovery Environment (XSEDE), which is supported by National Science Foundation grant number ACI-1548562.

**Reference:**

[1] Zhang Y, Tang T T, Girit C, Hao Z, Martin M C, Zettl A, Crommie M F, Shen Y R and Wang F 2009 *Nature* **459** 820.
[2] Ohta T, Bostwick A, Seyller T, Horn K and Rotenberg E. 2006 *Science* **313** 951-4.
[3] McCann E 2006 *Physical Review B*. **74** 161403.
[4] Castro E V, Novoselov K S, Morozov S V, Peres N M R, Dos Santos J L, Nilsson J, Guinea F, Geim A K and Neto A C 2007 *Physical review letters* **99** 216802.
[5] Johari P and Shenoy V B 2012 *ACS nano* **6** 5449-56.
[6] Conley H J, Wang B, Ziegler J I, Haglund Jr R F, Pantelides S T and Bolotin K I 2013 *Nano letters* **13** 3626-30.
[7] Lu N, Guo H, Li L, Dai J, Wang L, Mei W N, Wu X and Zeng X C 2014 *Nanoscale*, **6** 2879-86.
[8] Ramasubramaniam A, Naveh D and Towe E 2011 *Physical Review B* **84** 205325.
[9] Liu Q, Li L, Li Y, Gao Z, Chen Z and Lu J 2012 *The Journal of Physical Chemistry C* **116** 21556-62.
[10] Chu T, Ilatikhameneh H, Klimeck G, Rahman R and Chen Z 2015 *Nano letters* **15** 8000-7.
[11] Kuc A and Heine T 2015 *Chemical Society Reviews* **44**,2603-14.
[12] Deng B, Tran V, Xie Y, Jiang H, Li C, Guo Q, Wang X, Tian H, Koester S J, Wang H, Cha J J, Xia Q, Yang L and Xia F 2017 *Nature Communications* **8**.
[13] Li Y, Yang S and Li J 2014 *The Journal of Physical Chemistry C* **118** 23970-6.
[14] Liu Y, Qiu Z, Carvalho A, Bao Y, Xu H, Tan S J, Liu W, Castro Neto A H, Loh K P and Lu J 2017 *Nano Letters* **17** 1970-7.
[15] Kim J, Baik S S, Ryu S H, Sohn Y, Park S, Park B G, Denlinger J, Yi Y, Choi H J and Kim K S 2015 *Science* **349** 723-26.
[16] Dolui K and Quek S Y 2015 *Scientific reports* **5**.
[17] Miller D A, Chemla D S, Damen T C, Gossard A C, Wiegmann W, Wood T H and Burrus C A 1984 *Physical Review Letters* **53** 2173.
[18] Radisavljevic B, Radenovic A, Brivio J, Giacometti I V and Kis A 2011 *Nature nanotechnology* **6** 147-50
[19] Wang Q H, Kalantar-Zadeh K, Kis A, Coleman J N and Strano M S 2012 *Nature nanotechnology* **7** 699-712.
[20] Li L, Yu Y, Ye G J, Ge Q, Ou X, Wu H, Feng D, Chen X and Zhang Y 2014 *Nature nanotechnology* **9** 372-77.
[21] Sze Simon M and Kwok K Ng 2006 *Physics of semiconductor devices*. Hoboken, New Jersey John wiley & sons, 2006.
[22] Hohenberg P and Kohn W 1964 *Physical review* **136** B864.
[23] Resta R 1977 *Physical Review B* **16** 2717
[24] Perdew J P, Burke K and Ernzerhof M 1996 *Physical review letters* **77** 3865.
[25] Grimme S 2004 *Journal of computational chemistry* **25** 1463-73.
[26] Rydberg H, Dion M, Jacobson N, Schröder E, Hyldgaard P, Simak S I, Langreth D C and Lundqvist B I 2003 *Physical review letters* **91** 126402
[27] Wickramaratne D, Zahid F and Lake R K 2014 *The Journal of chemical physics* **140** 124710
[28] Ataca C, Sahin H and Ciraci S 2012 *The Journal of Physical Chemistry C* **116** 8983-99.
[29] Mak K F, Lee C, Hone J, Shan J and Heinz T F 2010 *Physical review letters* **105** 136805.
[30] Splendiani A, Sun L, Zhang Y, Li T, Kim J, Chim C Y, Galli G and Wang F 2010 *Nano letters* **10** 1271-5.
[31] Koo J, G S, Lee H and Yang L 2017 *Nanoscale*
[32] Sarma S D, Jalabert R and Yang S R E 1990 *Physical Review B* **41** 8288.
[33] Liang Y and Yang L 2015 Phys. Rev. Lett. **114** 063001.
[34] Gao S, Spataru C D and Yang L 2016 Nano Lett. **16** 5568.

**Figures**

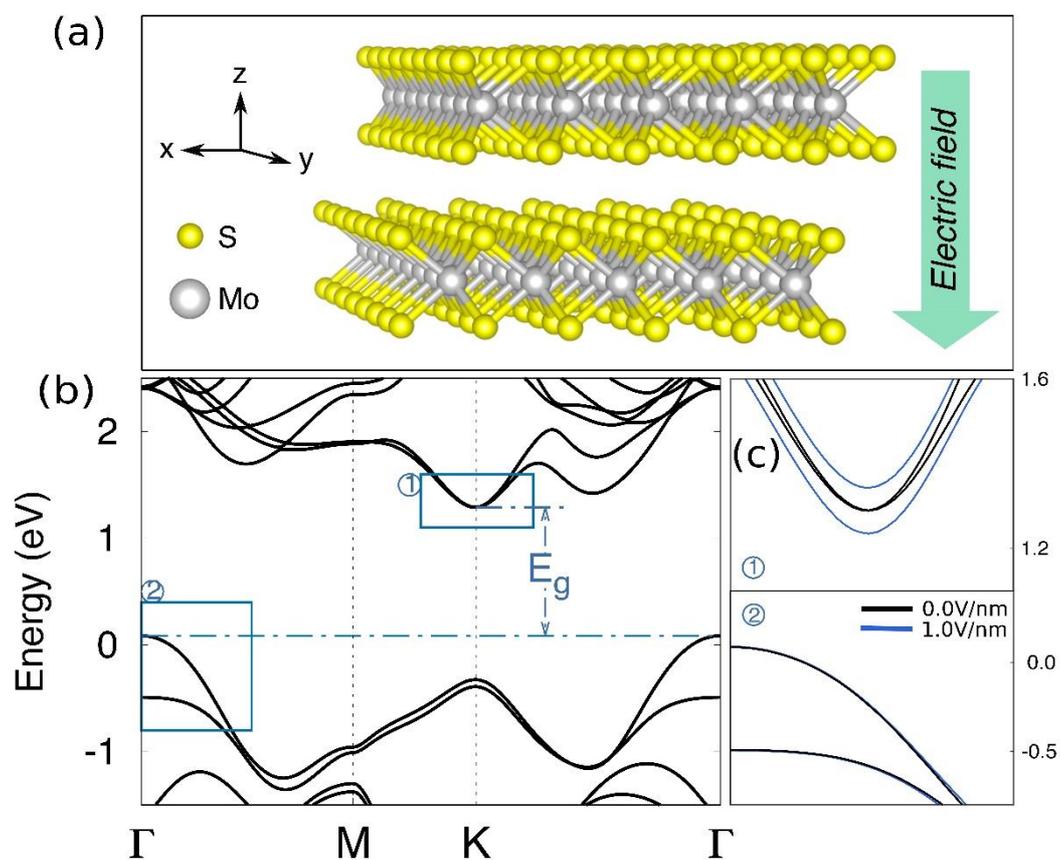

**Figure 1.** (a) The schematic ball-the-stick model of gated bilayer MoS$_2$. (b) the DFT calculated band structure of bilayer MoS$_2$. The top of valence band is set to be zero and the indirect band gap is marked. (c) The magnified band structures around the band edge (rectangles in (b)). The black lines are intrinsic band structures of bilayer MoS$_2$ while the blue lines are those of gated bilayer MoS$_2$.

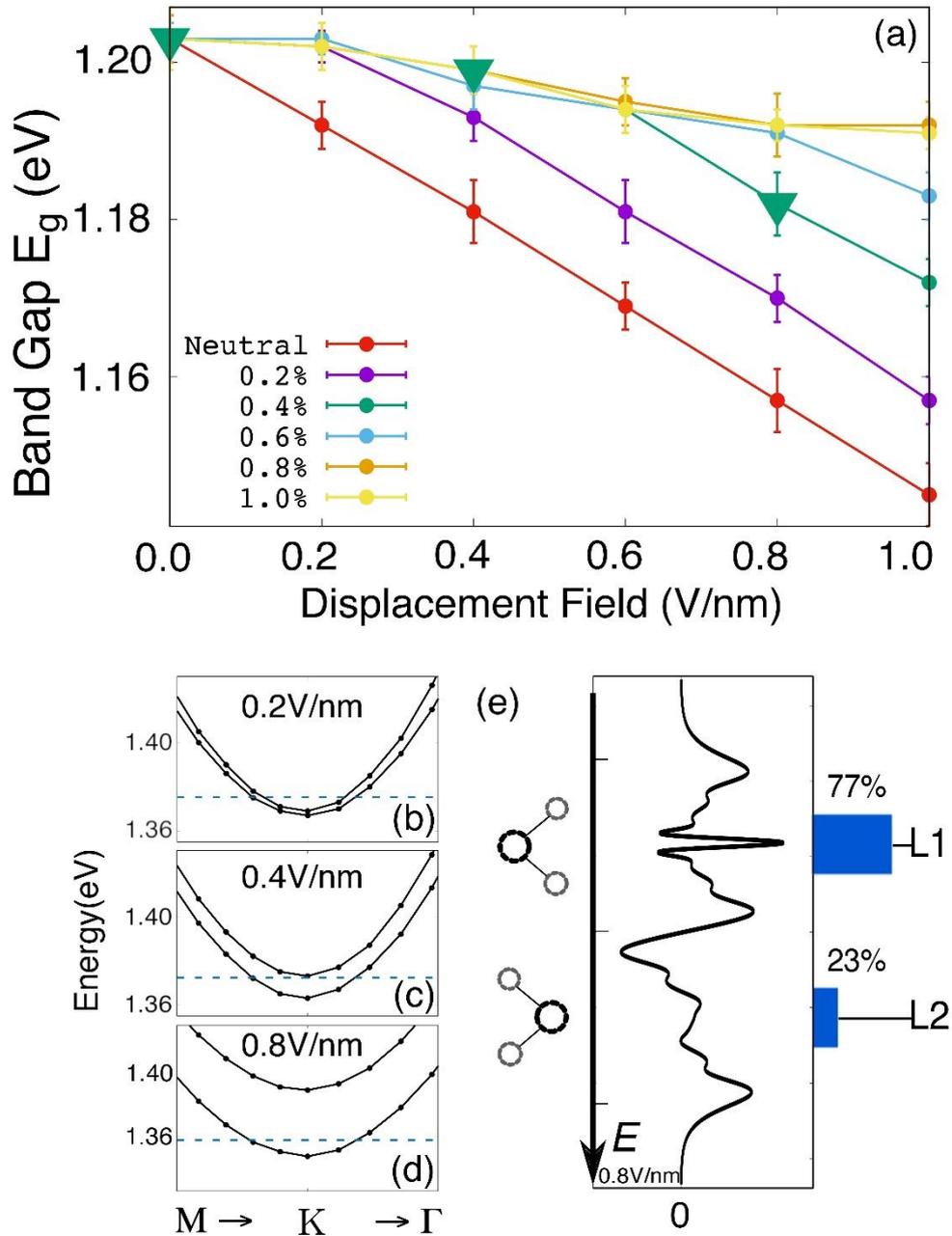

**Figure 2.** (a) The band gap of intrinsic and *n*-doped bilayer MoS$_2$ under gate field. (b), (c), and (d) are the magnified band structures around the conduction band edge. The dash line is the Fermi level. The corresponding band gaps are marked by triangles in (a). (e) The integrated carrier (electron) density under a 0.8V/nm gate field. The blue colored columns are the integrated carrier density in each layer (L1 and L2). The ball-and-the-stick model of the side view of bilayer is also MoS$_2$ plotted. The direction of applied gate field is marked as well.

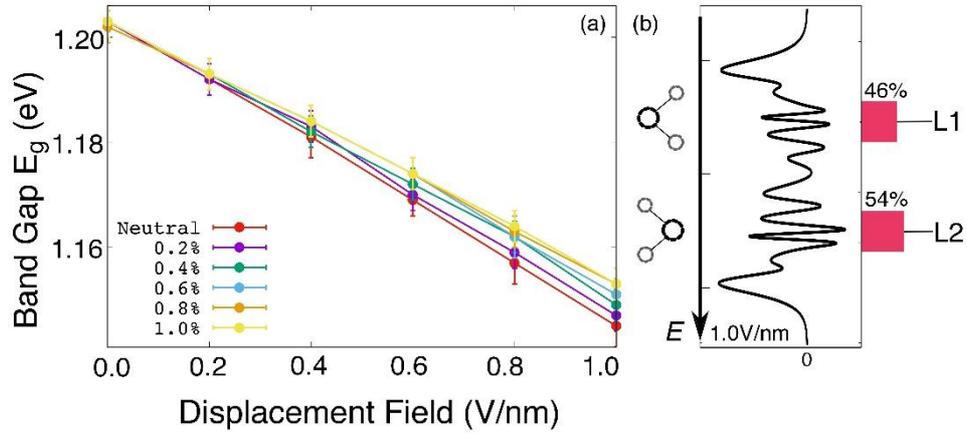

**Figure 3.** (a) The band gap of intrinsic and hole-doped (*p*-type) bilayer MoS$_2$ under gate field. (b) The integrated carrier (hole) density under a 1.0V/nm gate field. The red colored columns are the integrated carrier density in each layer (L1 and L2). The direction of applied gate field is marked as well.

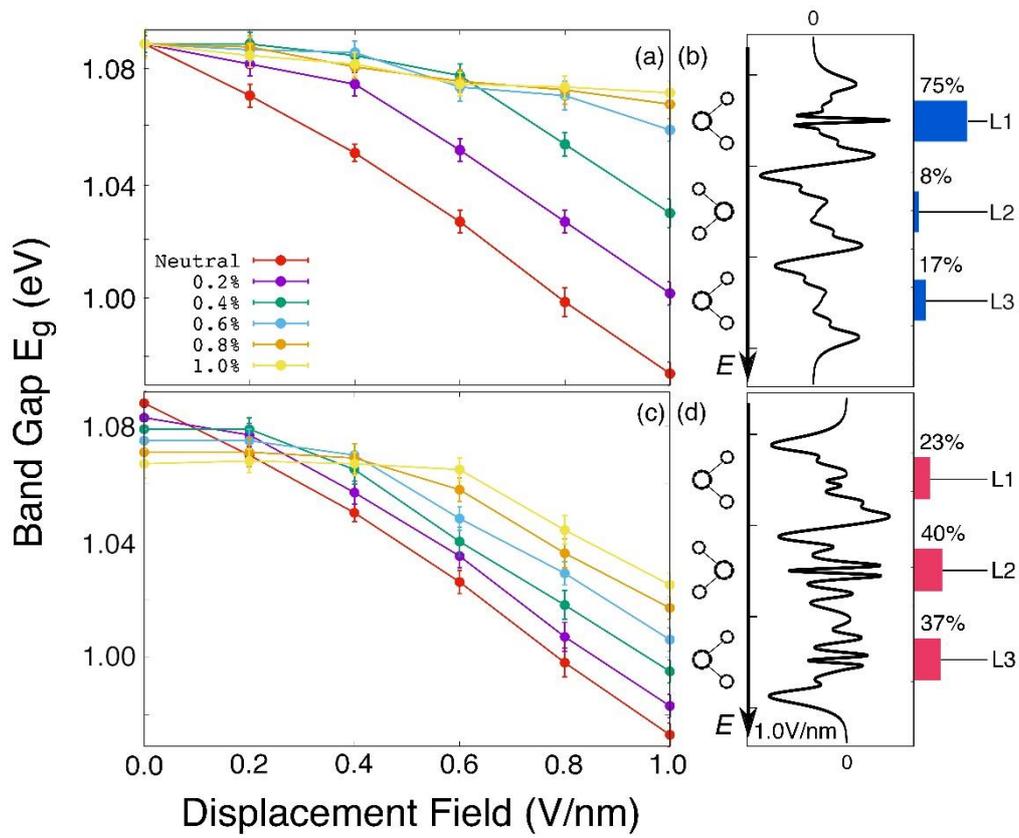

Figure 4. (a) The band gap of intrinsic and electron-doped trilayer MoS$_2$ under gate field. (b) The integrated carrier (electron) density under a 1.0-V/nm gate field. The blue-colored columns are the integrated carrier densities in each layer (L1, L2, and L3). (c) The band gap of intrinsic and hole-doped trilayer MoS$_2$ under gate field. (b) The integrated carrier (hole) density under a 1.0V/nm gate field. The red-colored columns are the integrated carrier densities in each layer (L1, L2, and L3). The direction of applied gate field is marked as well.